\pdfoutput=1

\documentclass[10pt,conference]{IEEEtran}
\IEEEoverridecommandlockouts

\usepackage{amsmath}
\usepackage{tikz}
\usepackage{filecontents}
\usepackage[labelfont=bf]{caption}
\usepackage{xcolor}
\usepackage{breakcites}
\usepackage{lineno}
\usepackage[linesnumbered,ruled,vlined,onelanguage]{algorithm2e}
\usepackage{hyphenat}
\usepackage{subcaption}
\usepackage{etoolbox}
\usepackage{verbatim}
\usepackage{pdfpages}

\usepackage[
backend=biber,
style=ieee,
sorting=none
]{biblatex}
\IEEEoverridecommandlockouts \IEEEpubid{\makebox[\columnwidth]{978-1-6654-0601-7/22\/\$31.00 ~\copyright~2022 IEEE \hfill} \hspace{\columnsep}\makebox[\columnwidth]{ }}

\newcommand\copyrighttext{%
  \footnotesize  This paper has been accepted for publication in IEEE/IFIP Network Operations and Management Symposium, Budapest, Hungary, April 2022. This is an author’s copy, The respective copyrights are with IEEE/IFIP.
}
\newcommand\copyrightnotice{%
\begin{tikzpicture}[remember picture,overlay]
\node[anchor=south,yshift=10pt] at (current page.south) {\fbox{\parbox{\dimexpr\textwidth-\fboxsep-\fboxrule\relax}{\copyrighttext}}};
\end{tikzpicture}%
}

\begin{document}

\title{E-GraphSAGE: A Graph Neural Network based\\ \ Intrusion Detection System for IoT}

% \author{\IEEEauthorblockN{Wai Weng Lo\IEEEauthorrefmark{1},
% Siamak Layeghy\IEEEauthorrefmark{1},
% Mohanad Sarhan\IEEEauthorrefmark{1}, 
% Marcus Gallagher\IEEEauthorrefmark{1}, and
% Marius Portmann\IEEEauthorrefmark{1}}

% \IEEEauthorblockA{\IEEEauthorrefmark{1}School of ITEE,
% University of Queensland, Brisbane QLD 4072, Australia }

\author{\IEEEauthorblockN{Wai Weng Lo\IEEEauthorrefmark{1},
Siamak Layeghy\IEEEauthorrefmark{2},
Mohanad Sarhan\IEEEauthorrefmark{3}, 
Marcus Gallagher\IEEEauthorrefmark{4}, and \\
Marius Portmann\IEEEauthorrefmark{5}}

\IEEEauthorblockA{School of Information Technology and Electrical Engineering\\
The University of Queensland, Brisbane, Australia}

\IEEEauthorblockA{Email: \IEEEauthorrefmark{1}w.w.lo@uq.net.au,
\IEEEauthorrefmark{2}siamak.layeghy@uq.net.au,
\IEEEauthorrefmark{3}m.sarhan@uq.net.au,
\IEEEauthorrefmark{4}marcusg@itee.uq.edu.au,
\IEEEauthorrefmark{5}marius@itee.uq.edu.au}

}


\maketitle

% \IEEEtitleabstractindextext{%
\begin{abstract}
This paper presents a new Network Intrusion Detection System (NIDS) based on Graph Neural Networks (GNNs). 
GNNs are a relatively new sub-field of deep neural networks, which can leverage the inherent structure of graph-based data. 
Training and evaluation data for NIDSs are typically represented as flow records, which can naturally be represented in a graph format. 
  In this paper, we propose \mbox{E-GraphSAGE}, a GNN approach that allows capturing both the edge features of a graph as well as the topological information for network intrusion detection in IoT networks.  To the best of our knowledge, our proposal is the first successful, practical, and extensively evaluated approach of applying GNNs on the problem of network intrusion detection for IoT using flow-based data. Our extensive experimental evaluation on four recent NIDS benchmark datasets shows that our approach outperforms the state-of-the-art in terms of key classification metrics, which demonstrates the potential of GNNs in network intrusion detection, and provides motivation for further research.
\end{abstract}

% }

% \maketitle

% \IEEEdisplaynontitleabstractindextext

\IEEEpeerreviewmaketitle

\copyrightnotice

%%%%%%%%%%%%%%%%%%%%%%%%%%%%%%%%%%%%%%%%%%%%%%%%%%%%%%%%%%%%%%
\section{Introduction}
IoT network attacks have significantly increased over the last few years, both in terms of frequency and level of sophistication.  IoT networks consist of many interconnected devices, including cameras, temperature sensors, smart TV, wireless printer, and other edge devices that require network connectivity \cite{ghasempour2019internet}. Cybercriminals can compromise and exploit IoT networks to perform malicious activities such as IoT ransomware, Botnet DDoS attacks. NIDSs, which are placed at strategic points within the IoT network to monitor traffic, are an important tool for the detection and mitigation of network-based cyber attacks. There are two main types of NIDS, signature-based and intrusion detection-based systems. A signature-based NIDS relies on a pre-installed set of attack signatures, which are compared and pattern-matched with the monitored network traffic in order to detect attacks. As a result, this type of NIDS can effectively detect known attacks with a relatively low false alarm rate, but they are significantly less effective in detecting new attacks or variants of existing attacks. In contrast, intrusion detection-based systems have the potential to detect intrusions by detecting deviations from the regular traffic patterns in the network. Over the last few years, there has been great progress in the application of new Machine Learning (ML) models and techniques, in particular deep learning-based approaches, for the development of new NIDS solutions. 

In this paper, we are exploring the use of GNNs, a relatively new sub-field of deep neural networks for IoT network intrusion detection. 
GNNs are tailored to applications with graph-structured data, such as social sciences, chemistry, and telecommunications, and are able to leverage the inherent structure of the graph data by building relational inductive biases into the deep learning architecture. This provides the ability to learn, reason, and generalise from the graph data, inspired by the concept of message propagation~\cite{gnn_suv}.

Network intrusion detection is typically performed on flow-based network data such as NetFlow \cite{Claise2004CiscoSN}, where flows are identified by the communication endpoints (IP address, L4 port number, L4 protocol), and annotated via a set of flow fields which provide details on the flows, such as the number of packets, number of bytes, flow duration, etc. 
This flow data can naturally be represented in a graph format, where the flow endpoints are mapped to graph nodes, and network traffic flows are mapped to the graph edges.  Both the topological information as well as the information contained in edge features are crucial for the classification of network traffic and the detection of attack flows.

Key recent works on ML-based NIDS, such as \cite{He}\cite{electronics9101565}\cite{Sarhan}\cite{ensmeble}\cite{s21020446},  consider flow data records independently, without considering the interrelation between them, and hence the traffic pattern more globally.
Consequently, these methods are limited in their ability to detect sophisticated IoT network attacks, such as botnet attacks  \cite{botnet}, distributed port scans \cite{port_scan}, or DNS Amplification attacks \cite{kambourakis2007detecting},  where a more global view of the network and traffic flow is required.
%

% Figure \ref{fig:DDoS} illustrates a botnet DDoS attack, where IoT devices are used as Zombies to perform a DDoS attack on a target. 

% For example, in a botnet DDoS attack, as \textcolor{blue}{illustrated in Figure \ref{fig:DDoS}}, attackers can use utilize IoT devices to perform DDoS attacks. 
% Most ML-based NIDSs  \cite{He}\cite{electronics9101565}\cite{Sarhan}\cite{ensmeble}\cite{s21020446} have not specifically considered the interrelated flow pattern.%s for the same reason 
% %(i.e. in this case, each of the IoT zombies to victim network flows). 
%
%------------------>

\begin{figure}[!t]
    \centering
        \includegraphics[width=0.85\columnwidth]{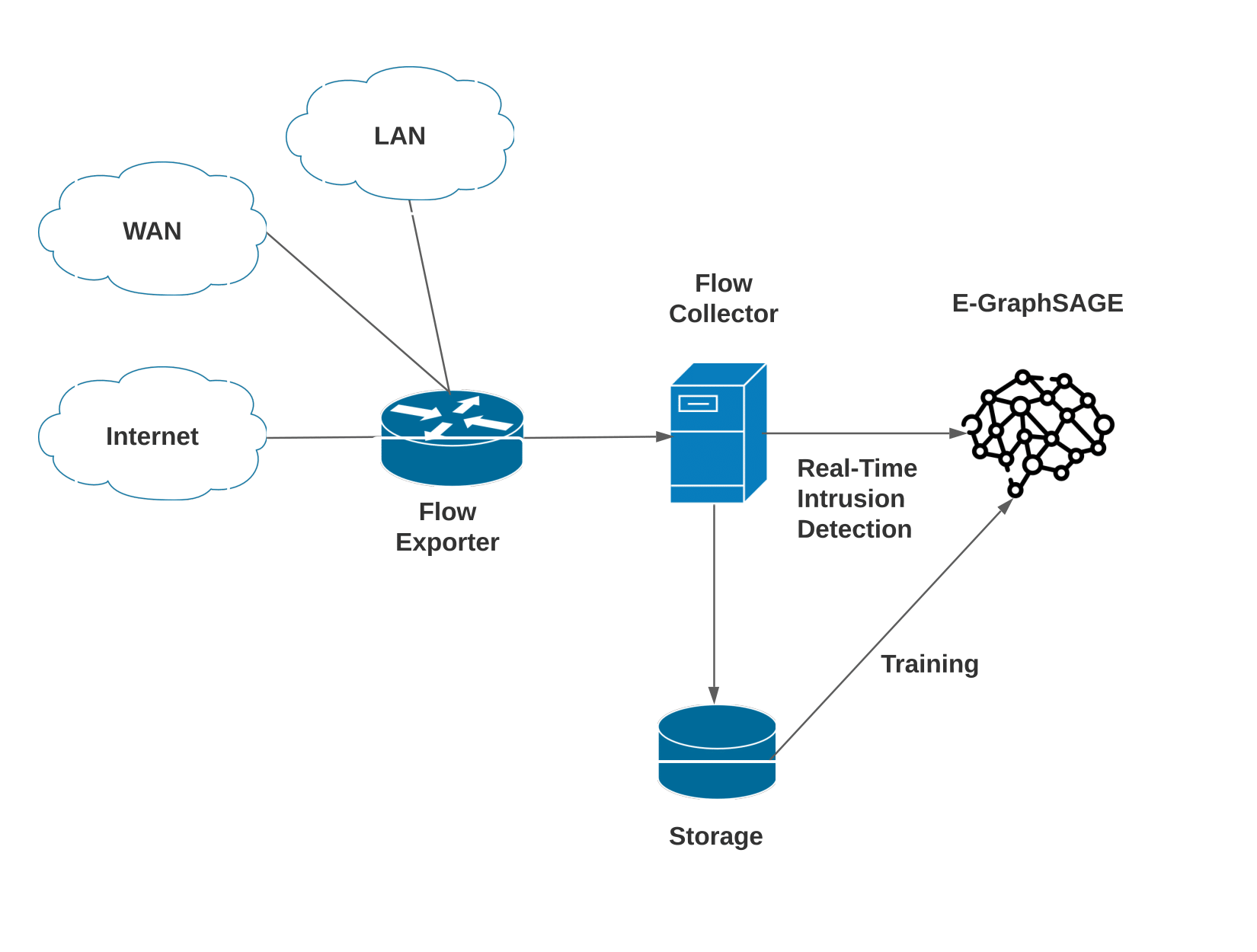}
        %\label{rfidtest_yaxis}
    \caption{Overall deployment architecture}
    \label{fig:Netflow_GNN}
\end{figure}

We believe that traffic patterns at a more global level, and hence flow interrelationships should be considered for the detection of distributed attacks in IoT, providing the motivation for the work presented in this paper. %
% , as the network flows are highly correlated for determining DDoS attack. Similar phenomena can also be found in distributed port scans and DNS amplification attacks.
%
We propose E-GraphSAGE, a GNN model that aims to overcome the current limitations of NIDSs for IoT. 
E-GraphSAGE enables the collection of information on flow-based features and topological patterns to account for the interconnected patterns of network flows.Consequently, E-GraphSAGE supports the process of edge classification, and hence the detection of malicious network flows, as illustrated in Figure \ref{fig:Netflow_GNN}.

We demonstrate how the E-GraphSAGE algorithm can be utilized to build a reliable NIDS, and provide an extensive experimental evaluation of the proposed system on four recent NIDS benchmark datasets.
Our results show that the E-GraphSAGE-based NIDS outperformed the state-of-the-art in regards to key classification metrics in all four considered benchmark datasets. 
To the best of our knowledge, our approach is the first successful, practical, and extensively evaluated approach of applying GNNs on the problem of network intrusion detection for IoT using flow-based data.

In summary, the key contributions of this paper are twofold:

\begin{itemize}
\item 
We have proposed and implemented E-GraphSAGE, a GNN-based NIDS, which allows the incorporation of edge features and topological patterns for IoT network intrusion detection. 

\item 

 We applied E-GraphSAGE on four benchmark IoT NIDS datasets for network intrusion detection, where the results demonstrated its potential via extensive experimental evaluation. 
\end{itemize}

The rest of the paper is organized as follows. Section~\ref{Related Work} discusses key related works, and Section~\ref{Background} provides the relevant background on GNNs and GraphSAGE.
Our proposed E-GraphSAGE algorithm and the corresponding NIDS are introduced in Section~\ref{Edge-Based GraphSAGE}. Experimental evaluation results are presented in Section~\ref{Experimental Results}, and Section \ref{Conclusions} concludes the paper. 

\section{Related Work}\label{Related Work}

Haitao et al.\cite{He} designed a multimodal sequential NIDS with a hierarchical progressive network to capture different levels of network data features. The approach is based on a combination of a deep autoencoder and LSTM architecture, in order to integrate the structural information within the temporal information shared between similar network connections. The system is evaluated on 3 datasets including the UNSW-NB15 dataset, where it achieved an accuracy of 96.8\% and 86.2\% in binary and multiclass classification experiments respectively.

In \cite{electronics9101565}, an attack mitigation framework for IoT networks is proposed using a hybrid of signature- and intrusion-based detection systems. The signature-based system uses a database of black-listed sources. The intrusion-based module adopts an extreme gradient boosting (XGBoost) algorithm. The XGBoost classifier achieved a binary classification accuracy of 99.99\% and a false alarm rate of 0.05\% on the BoT-IoT dataset, and 99.96\% multiclass classification F1-Score of  97\%.

Sarhan et al. \cite{Sarhan} converted the UNSW-NB15, BoT-IoT, and ToN-IoT datasets from their proprietary formats and feature set into a common Netflow-based format, with eight Netflow features as a common feature set. The corresponding new Netflow-based variants of the datasets, i.e. NF-UNSW-NB15, NF-BoT-IoT, and NF-ToN-IoT, have been made publicly available. 
The authors evaluated an Extra Tree ensemble classifier across these three datasets and have reported an F1-Score of 0.85, 0.97, and 1.00, respectively for binary classification. For the multi-class case, the corresponding F1-Scores are 0.98, 0.77 and 0.60.

In \cite{ensmeble}, the authors focused on securing Internet of medical things’ networks (IoMT) using a two-level intrusion detection model. The first level used a decision tree (DT), naive Bayes (NB), and random forest (RF) as first-level individual learners. In the next level, an XGBoost classifier was used to identify normal and attack instances. The ensemble model achieved a binary classification accuracy of 96.35\% and an F1-Score of 0.95 on the ToN-IoT dataset.

Churcher et al. \cite{s21020446} applied several machine learning algorithms such as k-nearest neighbour (KNN), decision tree (DT), support vector machine (SVM), naive Bayes (NB), random forest (RF), artificial neural network (ANN), and logistic regression (LR) for network intrusion detection.
The authors evaluated different classifier performances on BoT-IoT datasets and reported that the KNN classifier achieved the highest multiclass classification performance with an F1-Score of 0.99 and classification accuracy of 99.00\%.

Xiao et al. \cite{Xiao} proposed a graph embedding approach to perform intrusion detection on network flows.  The authors first converted the network flows into a first-order and second-order graph. The first-order graph learns the latent features from the perspective of a single host by using its IP address and port number.
The second-order graph aims to learn the latent features from a global perspective by using source IP addresses, source ports, destination IP addresses, as well as destination ports. 
The extracted graph embeddings and the raw features are then used to train a Random Forest classifier to detect network attacks.

However, a significant limitation of this approach is its use of a traditional transductive graph embedding method \cite{Hamilton2017}, which limits its ability to classify samples with graph nodes, e.g. IP addresses and port numbers, which were not seen during the training phase. This makes the approach unsuitable for most practical NIDS application scenarios. In contrast, the E-GraphSAGE approach presented in this paper uses an inductive learning approach, which does not suffer from this limitation.

% \subsection{Graph-Based NIDSs}
Zhou et al.\cite{Zhou} proposed using a graph convolutional neural network (GCN) to perform P2P botnet node detection. The authors first generated botnet traffic by creating botnet connections mixed with different real large-scale network traffic flows. Then, they applied the GCN for botnet node classification.The generated graph does not include any flow or node features and the approach only considers the topological information of the network connectivity graph for P2P botnet node classification, rather than flow classification. This approach limits the detection of botnet attacks, and does not focus on other network attacks, such as XSS and ransomware. In addition, it does not leverage all the information provided in network flow data, i.e., the network flow features.

While some existing graph representation learning methods have already considered edge features, none of them can be directly applied to network intrusion detection.
Methods such as  \cite{Gong_2019_CVPR} \cite{pmlr-v70-gilmer17a} consider edge features, but only for the purpose of improving node representation for better performance, and not edge classification, which is the aim in NIDSs. 

% In relational graph neural networks \cite{rgcn}, the authors consider modeling relational graphs with different relation types.

%-------->

% However the current benchmark datasets \cite{Sarhan} \cite{Koroniotis2019} \cite{ton-iot} have not held the heterogeneous nature across various network flows. 

% Others have focused on utilizing node features for message propagation  \cite{Gong_2019_CVPR} \cite{pmlr-v70-gilmer17a}. However, these methods cannot be applied to network intrusion detection because they are not designed for edge classification tasks. They only exploit edge features to improve node representations, rather than utilize them for graph edge classification. 
% %

% However the current benchmark datasets \cite{Sarhan} \cite{Koroniotis2019} \cite{ton-iot} have not held the heterogeneous nature across various network flows. Others have focused on utilizing node features for message propagation  \cite{Gong_2019_CVPR} \cite{pmlr-v70-gilmer17a}. However, these methods cannot be applied to network intrusion detection because they are not designed for edge classification tasks. They only exploit edge features to improve node representations, rather than utilize them for graph edge classification. 

In contrast to related studies, our approach leverages edge features that can be extracted from network flow data, and is critical for the detection of individual attack flows via its edge embedding approach. NIDSs based on traditional (non-graph-based) machine learning algorithms currently have not fully leveraged the structural and topological information inherent in network flow data for complex network attack detection, such as a botnet attack, which is one of the key strengths of GNNs. A key feature and novel contribution of E-GraphSAGE proposed in this paper, is its ability to leverage both, topological information and edge features in network flow data. Furthermore, related studies that apply a ML approach for NIDS typically use either, one, two, or a maximum of three NIDS benchmark datasets to evaluate the proposed systems. For the evaluation of the proposed GNN-based approach, we used four different benchmark datasets, which provide a higher degree of confidence in the robustness of the method and the ability to generalize to different types of network scenarios.

%%%%%%%%%%%%%%%%%%%%%%%%%%%%%%%%%%%%%%%%%%%%%%%%%%%%%%%%%%%%%%%%%%%
\section{Background}\label{Background}
\subsection{Graph Neural Networks (GNN)}
GNN is one of the most recent and fastest growing subareas in machine learning. Its power and potential lies in its ability to leverage the inherent graph structure of a lot of data encountered in real world application domains, such as social media networks, biology, telecommunications, etc. 
The graph format captures the structural information by modelling a set of objects and their relationships. The objects are represented by graph nodes and their relationships by graph edges. 
In the case of computer networks, individual hosts (IP addresses) are modelled as graph nodes, and the communication between the hosts, i.e., the network flows, are modelled as graph edges.

The key motivation of using a GNN for NIDSs is the ability to easily and directly exploit the rich structural information in the network flow data, which can be directly encoded in a graph format. 
While some traditional ML-based approaches attempt to work with graph data, this is typically quite cumbersome and relies on hand-engineered features.

A common task performed by GNNs is to generating node \textit{embeddings} \cite{Cai}, which aims to encode nodes as low-dimensional vectors, while maintaining their key relationships and graph position in the original format.  Node embedding is typically a key precursor to apply for downstream tasks such as node classification and node clustering \cite{Cai}. 
GNNs have recently received a lot of attention due to their convincing performance and high interpretability of the results through the visualisation of the node \textit{embeddings} \cite{Zhou2018GraphNN}.

\subsection{GraphSAGE}\label{GraphSAGE Algorithm}

The \textit{Graph SAmple and aggreGatE (GraphSAGE)} algorithm is one of the most well-known GNNs and was developed by Hamilton et al. \cite{Hamilton2017}~. 
In GraphSAGE, a fixed size sub-set is (uniformly randomly) sampled. This allows limiting the space and time complexity of the algorithm, irrespective of the graph structure (e.g. node degree
distribution) and batch size.

The GraphSAGE algorithm operates on a graph $\mathcal{G}(\mathcal{V}, \mathcal{E})$, where $\mathcal{V}$ is the set of nodes and $\mathcal{E}$ is the set of edges.
The node features of a node $v$ are represented as the vector $x_v$, and the complete set of node feature vectors  as $\{x_v, \forall v\in \mathcal{V}\}$.

A key hyperparameter of the GraphSAGE algorithm is the number of \textit{graph convolutional layers} $K$, which specifies the number of hops via which node information is aggregated at each iteration. 
Another important aspect  of GraphSAGE is the choice of a differentiable aggregator function $AGG_k, \forall k\in \{1,...,K\}$, to aggregate information from neighbor nodes. The algorithm is performed in both a forward and backward propagation stage, which we now discuss in more detail.

\subsubsection{Node Embedding}

% we can start like this:
% The main objective of the algorithm in the forward propagation is to generate node embeddings. ...
Similar to the convolution operation in convolutional neural networks (CNNs),  information relating to the local neighborhood of a node is collected and is used to compute the node embedding.
The GraphSAGE algorithm starts by assuming the model has already been trained and the weight matrices and aggregator function parameters are fixed. 
For each node, the algorithm iteratively aggregates information from the node's neighbours, the node's neighbours-neighbours, and so on. 

At each iteration, initially the neighborhood of the node is sampled, and the information from the sampled nodes is aggregated
into a single vector. 
At the $k$-th layer, the aggregated information $\mathbf{h}_{N(v)}^{k}$ at a node $v$, based of the sampled neighborhood $N(v)$, can be expressed as Equation \ref{eq:general_aggregator} \cite{Hamilton2017}:

\begin{equation}
\label{eq:general_aggregator}
    \mathbf{h}_{\mathcal{N}(v)}^{k} = \text { AGG }_{k}\left(\left\{\mathbf{h}_{u}^{k-1}, \forall {u} \in \mathcal{N}(v)\right\}\right)\
\end{equation}

Here, $\mathbf{h}_{u}^{k-1}$ represents the embedding of node $u$ in the previous layer. These embeddings of all nodes $u$ in the neighbourhood of $v$ are aggregated into the embedding of node $v$ at layer $k$.

\begin{figure}[!t]
    \centering
        \includegraphics[width=0.85\columnwidth]{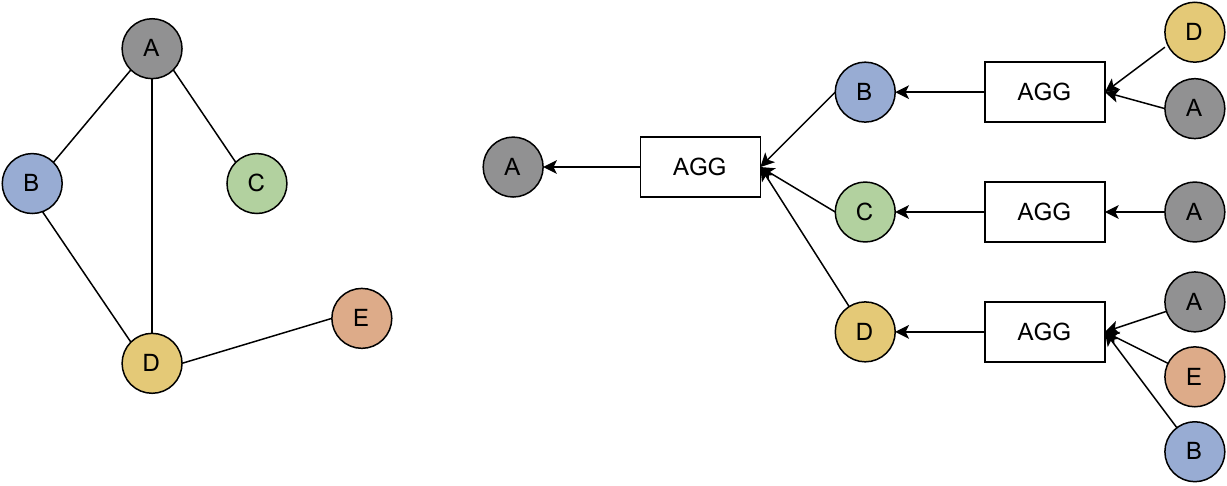}
        %\label{rfidtest_yaxis}
    \caption{A given graph (left), and the corresponding GraphSAGE architecture with depth-2 convolutions (right) and full neighborhood sampling.} 
    \label{fig:graphsage_overview}
\end{figure}

This aggregation process is illustrated in Figure \ref{fig:graphsage_overview} (right), the topological and node features in the graph, as collected and aggregated from the k-hop neighborhood of each graph node.
In the GraphSAGE model, we can use different aggregation methods, including mean, pooling, or different types of neural networks, e.g., Long Short-Term Memory (LSTM)~\cite{Hamilton2017}. 

The aggregated embeddings of the sampled neighborhood $\mathbf{h}_{N(v)}^{k}$ are then concatenated with the node's embedding from the previous layer $\mathbf{h}_v^{k-1}$. 
After applying the model's trainable parameters ($\mathbf{W}^k$, the trainable weight matrix) and passing the result through a non-linear activation function $\sigma$ (e.g. ReLU), the layer $k$ node $v$ embedding is calculated, as shown in Equation \ref{eq:general_graphsage} \cite{Hamilton2017}.
% e generalisation of the aggregation formula in GraphSAGE can be stated by Equation \ref{eq:general_graphsage} 

 \begin{equation}
 \small
 \label{eq:general_graphsage}
 \mathbf{h}_{v}^{k} = \sigma\left(\mathbf{W}^{k}\cdot\operatorname{CONCAT}\left(\mathbf{h}_{v}^{k-1}, \mathbf{h}_{\mathcal{N}(v)}^{k}\right)\right) 
\end{equation}

The final representation (embedding) of node $v$ is expressed as $\mathbf{z}_v$, which is essentially the embedding of the node at the final layer $K$, as shown in Equation \ref{eq:final_embedding}.
For the purpose of node classification, $\mathbf{z}_v$ can be passed through a sigmoidal neuron or softmax layer.

 \begin{equation}
 \label{eq:final_embedding}
     \mathbf{z}_v = \mathbf{h}_v^K,\ \ \ \ \ \forall v \in \mathcal{V}
 \end{equation}
%  where $K$ is the k-hop neighbour or number of graph convolutional layers. 
%  \vspace{0.5cm}
 
%  \item \textbf{Back Propagation}

\section{E-GraphSAGE}\label{Edge-Based GraphSAGE}

Traditional GNNs, e.g. GraphSAGE, have been successfully applied in a wide range of applications. However, these approaches mainly focus on node features for node classification, and are currently unable to consider edge features for edge classification. In contrast, our proposed \mbox{E-GraphSAGE} algorithm allows us to consider  the edge (flow) information in the embedding process. This provides the basis for computing the corresponding edge embeddings and enabling edge classification; that is, the classification of network flows into benign and attack flows. 

The goal of an NIDS is to detect and identify malicious traffic and flows. This corresponds to the problem of edge classification in our graph representation of flow datasets, where the key information is provided as edge features. 
Consequently, the current NIDS benchmark network datasets \cite{Sarhan} \cite{Koroniotis2019} \cite{ton-iot} provide network flow information as edge features, rather than node features, which only allows edge (flow) classification. In contrast, most related GNN research to date has focused on the problem of node classification, and the proposed solutions cannot be applied to the edge classification problem in the context of NIDS.

% Edge classification, considered in our approach, is quite distinct from the problem of node classification, on which most GNN works have focused on so far. 

% \textcolor{blue}{ In contrast, in our approach the problem is formulated as an edge classification problem because the current benchmark network attack datasets \cite{Sarhan} \cite{Koroniotis2019} \cite{ton-iot} only provide network flow information (edge features) rather than node features. Moreover, the node labels of the edge devices are not provided in the network attack datasets, which is why the problem cannot be formulated as a node classification problem.}

Our proposed approach (E-GraphSAGE) aims to overcome this limitation and facilitate the capture of edge features and topological information for intrusion detection.
This section introduces E-GraphSAGE in two parts. The first part discusses the E-GraphSAGE model and the extensions to the original GraphSAGE algorithm that were made to facilitate edge embedding and edge classification. In the second part, we discuss the application of E-GraphSAGE as an NIDS. 

% This section consists of two parts. In the first part, the \textit{Edge-based GraphSAGE  (E-GraphSAGE)} model, the method proposed in this paper for using the edge features with GraphSAGE algorithm is discussed. The next part of this section explains the proposed method for implementing an NIDS algorithm based on E-GraphSAGE.

\subsection{E-GraphSAGE Model}\label{E-GraphSAGE Model}
\subsubsection{Edge Embedding}

The message passing function in the original GraphSAGE algorithm only considers node features, and edge features are not taken into consideration~\cite{Hamilton2017}. 
To include the edge features, it is required to sample and aggregate the edge information of the graph. In addition, the final output of the algorithm needs to provide an edge embedding rather than the node embedding provided by the original algorithm. Our proposed E-GraphSAGE algorithm with these modifications is shown in Algorithm \ref{alg:E-GraphSAGE}.

%AAAAAAAAAAAAAAAAAAAAAAAAAAAAAAAAAAAAAAAAAAAA
\IncMargin{1.5em}
\begin{algorithm}[!t]\footnotesize
\SetKwData{This}{this}\SetKwData{Up}{up}
  \SetKwFunction{Union}{Union}\SetKwFunction{FindCompress}{FindCompress}
  \SetKwInOut{Input}{input}\SetKwInOut{Output}{output}

\Indp\Indpp
  \Input{
  Graph $\mathcal{G}(\mathcal{V}, \mathcal{E})$;\\ 
  input edge features $\left\{\mathbf{e}_{uv}, \forall {uv} \in \mathcal{E}\right\}$;\\
  input node features $\mathbf{x}_{v} = \{1, \ldots, 1\}$;\\ 
  depth $K$;\\ 
  weight matrices $\mathbf{W}^{k}, \forall k \in\{1, \ldots, K\}$;\\ non-linearity $\sigma$;\\ 
  differentiable aggregator functions ${AGG}_{k}$ ;
  \vspace{0.05cm}
  }
  
  \Output{
  Edge embeddings $\mathbf {z}_{uv}, \forall  {uv} \in \mathcal{E}$
    }
  \BlankLine
    $\mathbf{h}_{v}^{0} \leftarrow \mathbf{x}_{v}, \forall v \in \mathcal{V}$

  \For{$k\leftarrow 1\ \KwTo\ K$}{
    \For{$v \in \mathcal{V}$}{\label{forins}
    ${\mathbf{h}_{\mathcal{N}(v)}^{k} \leftarrow  \text { AGG}_{k}\left(\left\{\mathbf{e}_{uv}^{k-1}, \forall {u} \in \mathcal{N}(v),  uv \in \mathcal{E}\right\}\right) }$
     
     ${\mathbf{h}_{v}^{k}  \leftarrow \sigma\left(\mathbf{W}^{k}\cdot\operatorname{CONCAT}\left(\mathbf{h}_{v}^{k-1}, \mathbf{h}_{\mathcal{N}(v)}^{k}\right)\right) }$
     }
}
    $\mathbf{z}_v = \mathbf{h}_v^K$

  \For{$uv \in \mathcal{E}$}{
    ${\scriptstyle \mathbf{z}_{uv}^{K}  \leftarrow {CONCAT}\left(\mathbf{z}_{u}^{K}, \mathbf{z}_{v}^{K}\right)}$
}

\caption{E-GraphSAGE edge embedding}
\label{alg:E-GraphSAGE}
\Indp\Indpp
\end{algorithm}\DecMargin{1em}

%AAAAAAAAAAAAAAAAAAAAAAAAAAAAAAAAAAAAAAAAAAAA

Key differences to the original GraphSAGE algorithm \cite{Hamilton2017} are in regards to the algorithm input, the message passing/aggregator function, and the output. 

The input includes edge features $\left\{\mathbf{e}_{uv}, \forall {uv} \in \mathcal{E}\right\}$, which are, as mentioned before, missing from the list of inputs in GraphSAGE. 
Since the flow-based NIDS datasets only consist of flow (edge) features rather than node features. Therefore, we are using only provided edge features. We use the vector $\mathbf{x}_{v} = \{1, \ldots, 1\}$ to initialise the node features (and initial node embeddings) and the dimension of all one constant vector is the same as the number of edge features, as shown in Line 1 of the algorithm. 

In Line 4, instead of using the standard GraphSAGE node aggregator function (Equation \ref{eq:general_aggregator}), our newly proposed neighborhood aggregator function creates the \textit{aggregated embeddings of the sampled neighborhood edges} at the $k$-th layer, as shown in Equation~\ref{eq:general_edge_aggregator} below.

\begin{equation}
\small
\label{eq:general_edge_aggregator}
    \mathbf{h}_{\mathcal{N}(v)}^{k} = \text { AGG }_{k}\left(\left\{\mathbf{e}_{uv}^{k-1}, \forall {u} \in \mathcal{N}(v), uv \in \mathcal{E}\right\}\right)\
\end{equation}

Here, $\mathbf{e}_{uv}^{k-1}$ are the features of edge $uv$ from $\mathcal{N}(v)$, the sampled  neighborhood of node $v$, at layer \mbox{$k$-$1$}. 
The set $\{\forall {u} \in \mathcal{N}(v), uv \in \mathcal{E}\}$ represents the sampled edges in the neighborhood $\mathcal{N}(v)$). 
In Line 5, the node embedding for node $v$ at layer $k$ is calculated as in GraphSAGE (Equation \ref{eq:general_graphsage}), but with the critical difference that in the new  algorithm $\mathbf{h}_{\mathcal{N}(v)}^{k}$ is calculated via Equation \ref{eq:general_edge_aggregator} to include the edge features. Thus, the topological and edge information in the network flow graph is collected and aggregated from the k-hop neighborhood of each network graph node. 

The final node embeddings at depth $K$, $\mathbf{z}_v = \mathbf{h}_v^K$, are assigned in Line 8. Finally, the edge embeddings $\mathbf{z}_{uv}^{K}$ for each edge $uv$ are calculated as the concatenation of the node embeddings of nodes $u$ and $v$, as shown in Equation~\ref{eq:final_edge_embedding} below.

\begin{equation}
\small
\label{eq:final_edge_embedding}
\mathbf{z}_{uv}^{K} = {CONCAT}\left(\mathbf{z}_{u}^{K}, \mathbf{z}_{v}^{K}\right), uv \in \mathcal{E}
\end{equation}
This represents the final output of the forward propagation stage in E-GraphSAGE.

\subsubsection{Time and Space Complexity}\label{Computational Analysis}
The loops over the k-hop neighbour edges are the most time-consuming of the model as shown in Algorithm~\ref{alg:E-GraphSAGE}, Line 4. The upper bound time complexity estimation of E-GraphSAGE can be represented as $O\left(eKnd^{2}\right)$, where $n$ is the total number of nodes in the network, $e$ is the number of neighbour edges being sampled for each node, $K$ is the number of layers, $d$ is the dimension of the node hidden features. The space complexity of E-GraphSAGE is $O\left(beKd+K d^{2}\right)$, where $b$ is the batch size. Since E-GraphSAGE can support a min-batch setting, i.e., a fixed size of neighbour edges are being sampled to improve the training efficiency and reduce memory consumption. Instead of using full neighbourhood sets in Algorithm~\ref{alg:E-GraphSAGE},  the per-batch space and time complexity for E-GraphSAGE, which uniformly randomly samples a fixed-size set of edge neighbors, is $O\left(\prod_{i=1}^{K} E_{i}\right)$, where the sampled edges are $E_{i}, i \in\{1, \ldots, K\}$. 

% \end{itemize}
%%%%%%%%%%%%%%%%%%%%%%%%%%%%%%%%%%%%%%
% \subsection{Process}\label{Process}
\subsection{E-GraphSAGE NIDS}\label{E-GraphSAGE NIDS}
\begin{figure}[!t]
 \centering
\includegraphics[width=0.85\columnwidth]{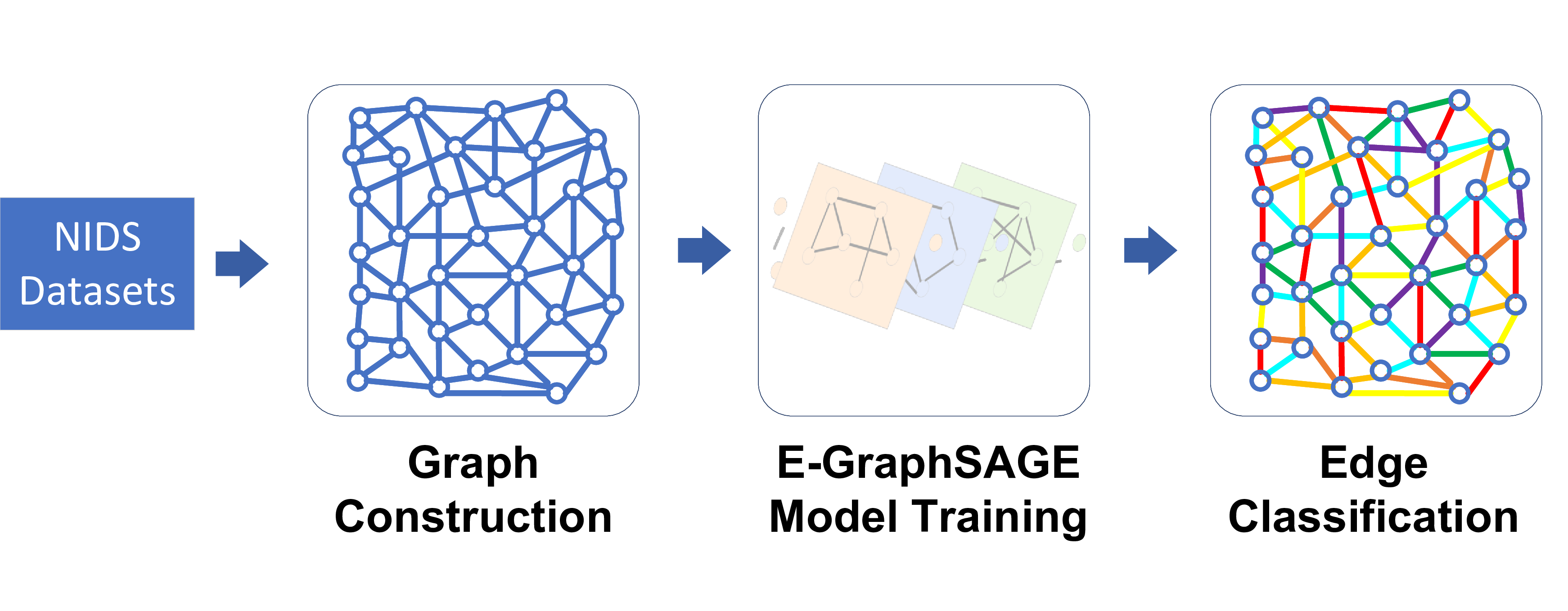}
\caption{Proposed E-GraphSAGE-based NIDS architecture.}
\label{fig:E-GraphSAGE NIDS}
\end{figure}

Figure~\ref{fig:E-GraphSAGE NIDS} shows a high level overview of our proposed NIDS based on E-GraphSAGE. 
%
% of our proposed E-GraphSAGE-based NIDS architecture. Although, there are data preprocessing stages such as data cleaning, formatting and standardisation, that is general in all ML-based NIDSs and have not been shown in the architecture. 
%
First, the network graph is generated from the network flow data, and is then fed into the supervised training process of the E-GraphSAGE model in the next step. In the last step, the edge embeddings are created, which forms the basis for the classification of edges (network flows) into benign and attack classes. These three steps are explained in the following subsections. 

%FFFFFFFFFFFFFFFFFFFFFFFFFFFFFFFFFFFFFFFFFFFFFFFFf
%FFFFFFFFFFFFFFFFFFFFFFFFFFFFFFFFFFFFFFFFFFFFFFFFf

\vspace{0.25cm}
\subsubsection{Network Graph Construction}\label{Network Graph Construction}
Network flows (e.g., NetFlow) are a common format for recording network communication in production networks, and it is also the most common format in the context of NIDS. The flow records usually consist of fields for identifying the source and destination of the communication, and the rest of a record field provides further information on the flow, such as the number of packets and bytes, flow duration, etc. 
A graph therefore provides  a very natural way to model such data.

% information relating the communications happening between the two ends. As such, the immediate way to model a flow into a graph is to model communication ends as graph nodes and the communications, as graph edges. 

There are various possibilities for defining the end points of a flow and hence the graph nodes.  The method used in this study uses the following 4 flow fields to identify a graph edge:  Source IP Address, Source (L4) Port, Destination IP Address, and Destination (L4) Port. The first two fields form a 2-tuple that identifies the source node and the two last fields identify the destination node of the flow. The additional flow fields provide features that are associated with the corresponding graph edge.  e.g. , the source node $(172.26.185.48 : 52962)$ exchanges data with destination node $(192.168.1.152 : 80)$ and the corresponding flow can be represented as a network edge.

To construct the network graph from the flow data, we mapped the original source IP addresses to randomly assigned IP addresses in the range from 172.16.0.1 to 172.31.0.1. The reason for this is the fact that in a lot of the NIDS datasets used in this paper, only a small number of IP addresses were used as the source of the attacks. The random mapping  avoids the potential problem of the source IP addresses providing an unintentional label for attack traffic.

% and the nodes on the graph are represented as computer nodes. 
Since in our graph construction all remaining flow record fields were assigned to the edge, the graph nodes are featureless.
% nodes with all one vector. 
%
In this algorithm, we assign a \textit{one vector}, a vector with all values of one to all nodes. As shown in Algorithm 1, we are using this method in our algorithm, with the dimensionality of the all one constant vector in each node depending on the number of edge features.

%%%%%%%%%%%%%%%%%%%%%%%%%%%%%%%%%%%%
\subsubsection{E-GraphSAGE Training}
The neural network model we use in our implementation consists of two \mbox{E-GraphSAGE} layers ($K=2$), which means that neighbour information is aggregated from a two-hop neighbourhood. 
For the aggregation function $AGG$, as shown in Equation \ref{eq:general_edge_aggregator}, we use the mean function which simply finds the element-wise mean of the edge features in the sampled neighborhood. The definition of the mean aggregator function in E-GraphSAGE is provided in Equation~\ref{eq:mean_edge_aggregator} below.

% The neural network model we use consists of two \mbox{E-GraphSAGE} layers ($K=2$, i.e., two-hop neighbours) that use the mean aggregation for the aggregator function (Equation \ref{eq:general_edge_aggregator}). The mean aggregator simply finds the element-wise mean of the edge embeddings in the sampled neighborhood 

\begin{equation}
\label{eq:mean_edge_aggregator}
    \mathbf{h}_{\mathcal{N}(v)}^{k} = 
    \sum_{
    \substack{
    {u} \in \mathcal{N}(v), \\
    uv \in \mathcal{E}
    }
    }\frac{\mathbf{e}_{uv}^{k-1}
    }{\rvert N(v)\rvert _{e}}
\end{equation}

Here,  $\rvert N(v)\rvert _{e}$ represents the number of  edges in the sampled neighborhood, and  $\mathbf{e}_{uv}^{k-1}$ represent their edge features at layer $k-1$. For our implementation, we chose full neighborhood sampling, which means information from the full set of edges in a node's neighbourhood is aggregated. 

In both E-GraphSAGE
% graph convolutional ?
layers, for the hidden feature sizes of each layer represented in Equation \ref{eq:general_graphsage}, we use a size of 128 hidden units, which is also the dimension of the node embeddings. 
For the nonlinear transformation ($\sigma$ in Equation \ref{eq:general_graphsage}),
we use the ReLU activation function, and for the purpose of regularisation, we use a dropout mechanism with a rate of $0.2$ between the two E-GraphSAGE layers. 
We chose the cross-entropy loss function, and gradient descent in the backward propagation phase is performed using the Adam optimizer  with learning rate of 0.001. 

When node embeddings are generated in the last \mbox{E-GraphSAGE} layer, they are converted to the corresponding edge embeddings (Line 10 of Algorithm \ref{alg:E-GraphSAGE}).
Since the edge embeddings are created via concatenation of two node embeddings, the size of the edge embeddings is 256 dimensions. 

The output edge embeddings pass through a softmax layer, which makes it possible to compare the output of the algorithm to the labels provided by the NIDS datasets and to tune the trainable model parameters in the backward propagation phase.
% (Section \ref{Backward Propagation} ).

\subsubsection{Edge Classification}
After the model parameters have been tuned in the training process, the model is ready for evaluation via the classification of unseen test samples. 
The test flow records are also converted to graphs, passed through the trained \mbox{E-GraphSAGE} layers, from which the edge embeddings are calculated. They are then converted to class probabilities in the final softmax layer and are finally compared with the true class labels in order to compute the classification evaluation performance metrics. 

%%%%%%%%%%%%%%%%%%%%%%%%%%%%%%%%%%%%%%%%%%%%%%%%%%%%%%%%%%%%%%%%%%%%%%%%%
\section{Datasets}\label{Datasets}

For the evaluation of the  GNN-based NIDS proposed in this paper, we use four publicly available NIDS datasets, which consist of different types of labelled attack flows as well as benign network flows. 
The first two datasets, consisting of ToN-IoT~\cite{ton-iot} and BoT-IoT~\cite{Koroniotis2019}, have proprietary formats and feature sets, and have been widely used to evaluate Machine Learning based network intrusion detection systems in IoT. In addition, we also consider two recently created variants of these datasets, where the respective format is translated to NetFlow with a common feature set, i.e. NF-TON-IoT \cite{Sarhan} and NF-BoT-IoT~\cite{Sarhan}. 
%These dataset files usually need to be processed to deal with missing, infinite and empty values. The non-numerical features were converted to numerical values using the label encoding technique. In addition, the standard scaler was used to remove the mean from the feature values and to scale them to unit variance. The flow identifiers in each dataset were removed to avoid learning bias towards attacking and victim endnodes. 
A brief overview of these datasets is provided in the following.

\begin{itemize}
\item \textbf{BoT-IoT:} This is another recent dataset with a specific on IoT networks, created by Koroniotis et al. \cite{Koroniotis2019} in 2019. 
The authors simulated IoT services such as water stations, by using the Node-red tool \cite{Node-RED} and generated the corresponding IoT traffic. The Argus tool was consequently used for feature extraction. This dataset is comprised of 6 types of attacks and a total of 47 features with corresponding class labels. The dataset contains only 477 (0.01\%) benign flows and 3,668,045 (99.99\%) attack flows, with a total of 3,668,522 flows.

\item \textbf{ToN-IoT:} This is a relatively new and extensive dataset that was generated in 2019 by Abdullah et al.\cite{ton-iot}, which includes different types of IoT data, such as operation system logs, telemetry data of IoT/IIoT services, as well as IoT network traffic collected from a medium-scale network at the Cyber Range and IoT Labs at the UNSW Canberra (Australia). 
In this paper, only the network traffic component of the dataset is used.
% %I am not sure what this sentence mean
% The paper indicated that a few of the network attack datasets only concentrate on IoT and IIoT traffic; Therefore, the cureent network attack datasets did not reflect the current IoT and IIoT trends. 
%
The Bro-IDS network monitoring tool was used to generate the dataset's 44 network flow features. 
The dataset consists of 796,380 (3.56\%) benign flows and 21,542,641 (96.44\%) attack flows, with a total of 22,339,021 flows.

\item \textbf{NF-ToT-IoT and NF-BoT-IoT:} A lack of a standard format and feature set among the various NIDS datasets has made it very difficult to compare the performance of ML-based network traffic classifiers across different datasets, and to evaluate their ability to generalise to different network scenarios. 
Sarhan et al.~\cite{Sarhan} have addressed this problem by providing the NetFlow version of the above mentioned three NIDS datasets. The authors of ~\cite{Sarhan} used the raw packet capture (pcap) files of the original NIDS datasets and converted them to the NetFlow format via the nprobe \cite{nprob} tool and selected 12 fields to be extracted, resulting in the new variants of the original datasets, i.e. NF-ToT-IoT and NF-BoT-IoT. 
In NF-ToT-IoT, the total number of network flows is 1,379,274, out of which 1,108,995 (80.4 \%) are attack flows and 270,279 (19.6 \%) are benign flows. The NF-BoT-IoT dataset has a total number of 600,100  flows, out of which 586,241 (97.69 \%) are attack flows and 13,859 (2.31 \%) are benign flows. 

\end{itemize}
%\vspace{0.5cm}

%%%%%%%%%%%%%%%%%%%%%%%%%%%%%%%%%%%%%%%%%%%%%%%%%%%%%%%%%%%%%%%%%%%%%%%%%

\section{Experimental Results}\label{Experimental Results}
To evaluate the performance of the proposed neural network model, the standard metrics listed in Table~\ref{tab: metrics} are used, where $TP$, $TN$, $FP$ and $FN$ represent the number of True Positives, True Negatives, False Positives and False Negatives respectively.
First, the results of the binary classification problem, where the aim is to distinguish between attack and benign traffic, are presented. Subsequently, we present the results of the multiclass classification experiments, where the aim was to identify the specific attack type of each flow.

%TTTTTTTTTTTTTTTTTTTTTTTTTTTTTTTTTTTTTTTTTTT
\begin{table}[!b]\small
\renewcommand{\arraystretch}{1.4}
\caption{Evaluation metrics utilised in this study}
\centering
\begin{tabular}
{|c|c|} \hline
\textbf{Metric} & \textbf{Definition} \\ \hline 
\small{ Detection Rate (Recall)} &  $\frac{TP}{TP+FN}$ \\ \hline
\small{Precision} & $\frac{TP}{TP+FP}$ \\ \hline
\small{F1-Score} & $ 2 \times \frac{Recall\times Precision}{Recall + Precision}$ \\ \hline
\small{Accuracy} &   $\frac{TP+TN}{TP+FP+TN+FN}$ \\ \hline
\small{False Alarm Rate (FAR)} &   $\frac{FP}{FP+TN}$ \\ \hline

\end{tabular}
\label{tab: metrics}
\end{table}
%TTTTTTTTTTTTTTTTTTTTTTTTTTTTTTTTTTTTTTT

% TTTTTTTTTTTTTTTTTTTTTTTTTTTTTTTTTTTTTTTT
\begin{center}
\begin{table}[!t]
\caption{E-GraphSAGE binary classification results}
\resizebox{\columnwidth}{!}{\begin{tabular}{|c||c|c|c|c|c|c|c}
\hline \textbf{Dataset} & \textbf{Accuracy} & \textbf{Precision} & \textbf{F1-Score} & \textbf{Recall (DR)} & \textbf{FAR} %& Prediction Time (µs) 
\\
\hline BoT-IoT & 99.99\% & 1.00 & 1.00 & 99.99\% & 0.00 \% %& 0.24 
\\
NF-BoT-IoT &  93.57\% & 1.00 & 0.97 & 93.43\%  & 0.38 \% %& 0.16
\\
\hline ToN-IoT & 97.87\% & 1.00 & 0.99 & 97.86\% &  1.92\%  %& 0.14 
\\
NF-ToN-IoT & 99.69\% & 1.00 & 1.00 & 99.85\% & 0.15\%  %& 0.14
\\
\hline
\end{tabular}
}
\label{tab:binary_classification}
\end{table}
\end{center}
%TTTTTTTTTTTTTTTTTTTTTTTTTTTTTTTTTTTTTTTTT

%TTTTTTTTTTTTTTTTTTTTTTTTTTTTTTTTTTTTTTTTTTTT
\begin{table}[!b]\small
\centering
\caption{Performance of binary classification by E-GraphSAGE compared with the state-of-art algorithms}
\begin{tabular}{|c||c|c|c|}
%  \hline
%  \multicolumn{4}{|c|}{Binary Classification Performance Comparison with State-of-Arts Algorithms} \\
 \hline
  \textbf{Method} & \textbf{Dataset}  & \textbf{F1}\\

\hline
\textbf{  Proposed E-GraphSAGE }  & BoT-IoT   &   \textbf{1.00} \\
  XGBoost  \cite{electronics9101565}  & BoT-IoT  &   0.99\\
  \hline

\textbf{ Proposed E-GraphSAGE }  & NF-BoT-IoT    &   \textbf{0.97} \\
  Extra Tree Classifier \cite{Sarhan}   & NF-BoT-IoT   &  0.97 \\
   \hline

\textbf{  Proposed E-GraphSAGE }  & ToN-IoT   &   \textbf{0.99}\\
    Ensemble \cite{ensmeble}  & ToN-IoT   &   0.95 \\
      \hline

\textbf{   Proposed E-GraphSAGE}  & NF-ToN-IoT   &    \textbf{1.00} \\
  Extra Tree Classifier \cite{Sarhan}   & NF-ToN-IoT   &   1.00 \\
 \hline
\end{tabular}
\label{table:binary_comp}
\end{table}
%TTTTTTTTTTTTTTTTTTTTTTTTTTTTTTTTTTTTTTTTTTTT

\vspace{-5mm}
\subsection{Binary Classification Results}\label{Binary Classification Results}
The datasets used in this paper all include two sets of labels. The first label determines if a flow record belongs to a benign or attack class, and the second label identifies the attack type. We use the first label set for binary classification, and the second set for multiclass classification.

%FFFFFFFFFFFFFFFFFFFFFFFFFFFFFFFFFFFFFFFFFFFFFFFFFFFF

% FFFFFFFFFFFFFFFFFFFFFFFFFFFFFFFFFFFFFFFFFFFFFFF

We use the full dataset, except for the ToN-IoT dataset, where we considered a randomly sampled subset with 10\% of the original size, due to the very large size of the original dataset. 
In regards to training and evaluation data split, 70\% of the flow records of each datasets were selected for training and 30\% were reserved for testing and evaluation.

As mentioned above, the experimental evaluation was based on three original datasets with proprietary flow formats and feature sets, as well as their corresponding NetFlow versions. 
As shown in \cite{Sarhan}, even though the original dataset and its NetFlow counterpart represent essentially the same network events, the  performance of a classifier for the two datasets versions can vary significantly, due to the different feature sets.  
Table~\ref{tab:binary_classification} presents the detailed results of our \mbox{E-GraphSAGE} classifier for the binary classification case, showing Accuracy, Precision , F1-Score, Recall and FAR, for the considered four benchmark datasets. 
As can be seen in the table, the E-GraphSAGE classifier performs very well across all different performance metrics. Since the considered datasets are generally highly imbalanced, the F1-Score is a more relevant performance metric. 
We use the F1-Score to compare our classifier with the state-of-the-art, i.e., the best classification results reported in the literature for each of the four NIDS datasets. 

Table \ref{table:binary_comp} shows the corresponding results of our \mbox{E-GraphSAGE} classifier compared to the best performing related works for each datasets in terms of F1-Score. 
As can be observed from the table, in regards to F1-Score, E-GraphSAGE outperforms the best reported classifiers in ToN-IoT and BoT-IoT. The NF-ToN-IoT and NF-BoT-IoT experiments achieved an F1-score of 1.0 and 0.97, respectively,  which are the same as the state-of-art algorithms.

These results demonstrate the power of our GNN-based approach for traffic classification and network intrusion detection. In contrast to most related works, where a new ML-based NIDS classifier is evaluated on one, two or a maximum three datasets, we have demonstrated the performance of our E-GraphSAGE approach across four significantly different NIDS benchmark datasets. The consistently solid results demonstrate the robustness of our GNN-based approach and its ability to generalise across different types of network traffic patterns, feature sets, and attack types. We believe this is due to the ability of the GNN architecture to capture the information in the inherent graph structure of the flow-based NIDS datasets. 

%%%%%%%%%%%%%%%%%%%%%%%%%%%%%%%%%%%%%%%%
\subsection{Multiclass Classification Results}\label{Multiclass Classification Results}

%TTTTTTTTTTTTTTTTTTTTTTTTTTTTTTTTTTTTTTTTTT 
We are now considering the multiclass classification problem, where the classifier aims to distinguish between different types of attacks as well as benign traffic. This is a much harder problem than in the binary case. 
For the evaluation of the E-GraphSAGE classifier in the multiclass scenario, we considered the same four NIDS datasets as for the binary case. 
Depending on the NIDS dataset, the number of attack classes range between 4 and 9.  
%

%TTTTTTTTTTTTTTTTTTTTTTTTTTTTTTTTTTTTTTTTTT
\begin{table}[!t]\small
\centering
\caption{Results of multiclass classification by E-GraphSAGE on BoT-IoT dataset families}
\label{tab:multi-bot-iot}
\begin{tabular}{l|r|r|r|r|}
\cline{2-5}
                                               & \multicolumn{2}{c|}{\textbf{BoT-IoT}} & \multicolumn{2}{c|}{\textbf{NF-BoT-IoT}} \\ \hline
\multicolumn{1}{|c|}{\textbf{Class Name}} &
  \multicolumn{1}{c|}{\textbf{DR}} &
  \multicolumn{1}{c|}{\textbf{F1-Score}} &
  \multicolumn{1}{c|}{\textbf{DR}} &
  \multicolumn{1}{c|}{\textbf{F1-Score}} \\ \hline
\multicolumn{1}{|l|}{{Benign}}          & 100.00\%              & 0.99             & 99.45\%
              & 0.42            \\ \hline
\multicolumn{1}{|l|}{{DDoS}}        & 99.99\%               & 1.00            & 40.82\%
              & 0.39              \\ \hline
\multicolumn{1}{|l|}{{DoS}}        & 99.99\%              & 1.00             & 57.13\%   & 0.47              \\ \hline
\multicolumn{1}{|l|}{{Reconnaissance}}             & 99.98\%              & 1.00             & 84.50\%
             & 0.92              \\ \hline
\multicolumn{1}{|l|}{{Theft}}        & 93.75\%              & 0.97           & 99.83\%
              & 0.39              \\ \hline

% \multicolumn{1}{|l|}{\textbf{Accuracy}}        & \multicolumn{2}{c|}{\textbf{98.19\%}}   & \multicolumn{2}{c|}{\textbf{97.62\%}}      \\ \hline
\multicolumn{1}{|l|}{\textbf{Weighted Average}}         & \textbf{99.99\%}              & \textbf{1.00}             & \textbf{78.16\%}               & \textbf{0.81}               \\ \hline
% \multicolumn{1}{|l|}{\textbf{Prediction Time (\textmu s)}} & \multicolumn{2}{c|}{\textbf{0.51}}      & \multicolumn{2}{c|}{\textbf{0.14}}         \\ \hline
\end{tabular}
\end{table}
%TTTTTTTTTTTTTTTTTTTTTTTTTTTTTTTTTTTTTTTTTT

%TTTTTTTTTTTTTTTTTTTTTTTTTTTTTTTTTTTTTTTTTT
\begin{table}[!b]\small
\centering
\caption{Results of multiclass classification by E-GraphSAGE on ToN-IoT datasets families}
\label{tab:multi-ton-iot}
\begin{tabular}{l|r|r|r|r|}
\cline{2-5}
                                               & \multicolumn{2}{c|}{\textbf{ToN-IoT}} & \multicolumn{2}{c|}{\textbf{NF-ToN-IoT}} \\ \hline
\multicolumn{1}{|c|}{\textbf{Class Name}} &
  \multicolumn{1}{c|}{\textbf{DR}} &
  \multicolumn{1}{c|}{\textbf{F1-Score}} &
  \multicolumn{1}{c|}{\textbf{DR}} &
  \multicolumn{1}{c|}{\textbf{F1-Score}} \\ \hline
\multicolumn{1}{|l|}{{Benign}}          & 88.12\%              & 0.91             & 98.86\%
              & 0.92            \\ \hline
\multicolumn{1}{|l|}{{Backdoor}}        & 5.06\%               & 0.08            & 98.38\%
              & 0.99              \\ \hline
\multicolumn{1}{|l|}{{DDoS}}        & 96.94\%              & 0.98             & 52.35\%   & 0.68              \\ \hline
\multicolumn{1}{|l|}{{DoS}}             & 96.08\%              & 0.73            & 0.00\%
             & 0.00             \\ \hline
\multicolumn{1}{|l|}{{Injection}}        & 88.94\%              & 0.83           & 93.15\%
              & 0.71              \\ \hline
\multicolumn{1}{|l|}{{MIMT}}        & 87.43\%             & 0.18           & 22.88\%
              & 0.28              \\ \hline
\multicolumn{1}{|l|}{{Ransomware}}        & 98.55\%            & 0.94           & 96.49\%
              & 0.23              \\ \hline
\multicolumn{1}{|l|}{{Password}}        & 89.15\%             & 0.91           & 19.92\%
              & 0.25            \\ \hline
\multicolumn{1}{|l|}{{Scanning}}        & 75.84\%             & 0.85           & 15.32\%
              & 0.13              \\ \hline
\multicolumn{1}{|l|}{{XSS}}        & 92.08\%
            & 0.95           & 0.00\%
              & 0.00              \\ \hline

% \multicolumn{1}{|l|}{\textbf{Accuracy}}        & \multicolumn{2}{c|}{\textbf{98.19\%}}   & \multicolumn{2}{c|}{\textbf{97.62\%}}      \\ \hline
\multicolumn{1}{|l|}{\textbf{Weighted Average}}         & \textbf{86.78\%}              & \textbf{0.87}             & \textbf{67.16\%}               & \textbf{0.63}               \\ \hline
% \multicolumn{1}{|l|}{\textbf{Prediction Time (\textmu s)}} & \multicolumn{2}{c|}{\textbf{0.14}}      & \multicolumn{2}{c|}{\textbf{0.21}}         \\ \hline
\end{tabular}
\end{table}
%TTTTTTTTTTTTTTTTTTTTTTTTTTTTTTTTTTTTTTTTTTTTTTTTTTTTTT

Tables \ref{tab:multi-bot-iot} and \ref{tab:multi-ton-iot} show the corresponding results for the BoT-IoT and ToN-IoT datasets and their respective Netflow counterparts.
For the BoT-IoT dataset with its original format and feature set, a very high weighted Detection Rate and F1-Score is achieved for all the 5 traffic classes (4 attack classes plus benign class), with a weighted average of 99.99\% and 1.00 respectively. The corresponding numbers for the NF-BoT-IoT dataset are significantly lower, with a weighted average DR value of 78.16\% and weighted F1-Score of 0.81. 

The results for the ToN-IoT datasets show a high weighted average Detection Rate (86.78\%) and F1-Score (0.87), but the values vary significantly for the individual traffic classes. 
As for the BoT-IoT dataset, we observe a significantly lower classification performance for the NetFlow variant of the BoT-IoT dataset. This can be explained by the very different nature of the feature set. 
While the original feature sets include specifically engineered features to detect the attack classes included in the dataset, the NetFlow variants consist of relatively simple and generic features.

% The results of classification related to each pair of datasets, the dataset in proprietary format and its NetFlow format (NF) counterpart, are shown in Tables~\ref{tab:multi-unsw-nb15}, \ref{tab:multi-bot-iot} and \ref{tab:multi-ton-iot} for UNSW-NB15, BoT-IoT and ToN-IoT (and their NF counterparts) respectively.
% % 
% As seen, the overall classification performance in terms of the weighted average of Detection Rate (DR) and F1-Score is very high in three out of four datasets. 
% %
% % In the case of the BoT-IOT dataset, the weighted average of DR and F1-Score are 99\% and 100\% respectively.
% In three out of four datasets, the performance of E-GraphSAGE is not similarly high.

%

\begin{figure}[!t]
    
    \centering
 \subfloat[\centering ][raw data]
        {\includegraphics[width=0.50\columnwidth]{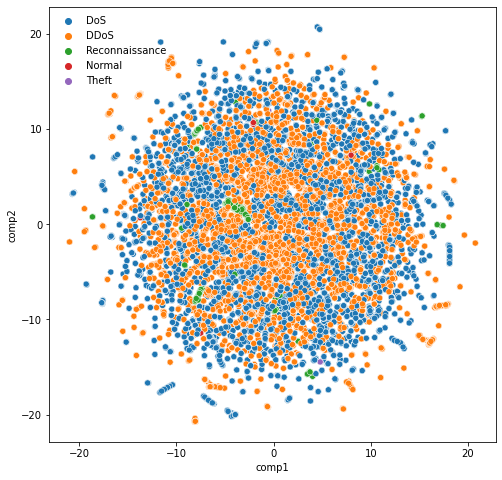}}
    \centering
    \subfloat[\centering ][edge embedded data]
    {\includegraphics[width=0.50\columnwidth]{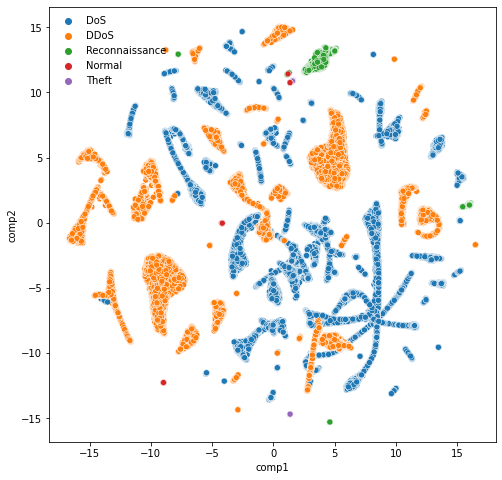}}
    \caption{Visualisation of dimensionality reduction a) Sample of BoT-IoT raw validation data, b)Sample of edge embeddings generated by E-GraphSAGE (Multiclass). }
    \label{fig: visualisation-multi}
\end{figure}

To get an intuitive understanding of the promising and robust performance of the E-GraphSAGE classifier, we investigated and visualised the edge embeddings generated by the algorithm, based on the BoT-IoT dataset.
For this, we generate two visualisations. First, we take the sample `raw' BoT-IoT dataset and the edge embedding then applied the \textit{Uniform Manifold Approximation and Projection (UMAP)} \cite{umap} dimensionality reduction algorithm to map the high-dimensional data down to two dimensions, for the purpose of visualisation. 
The result is shown in Figure~\ref{fig: visualisation-multi}. We now see a clear separation of attack flows and normal/benign flows. This illustrates the ability of the edge embedding mechanism of the E-GraphSAGE algorithm to leverage the inherent graph structure of the flow-based NIDS data, and to be able to separate attack traffic from normal traffic in the embedding space, which in turn results in a high classification performance. 

%FFFFFFFFFFFFFFFFFFFFFFFFFFFFFFFFFFFFFFFFFFFF
%FFFFFFFFFFFFFFFFFFFFFFFFFFFFFFFFFFFFFFFFFFFF

%TTTTTTTTTTTTTTTTTTTTTTTTTTTTTTTTTTTTTTTTTTTTTTTTTTTTTTT
\begin{table}[!b]\small
\centering
\caption{Performance of multiclass classification by E-GraphSAGE compared with the state-of-art algorithms}
\begin{tabular}{|c||c|c|}
%  \hline
%  \multicolumn{4}{|c|}{Multiclass Classification Performance Comparison with State-of-Arts Algorithms} \\
 \hline
  \textbf{Method} & \textbf{Dataset} & \textbf{W-F1}\\
  \hline
 \textbf{ Proposed E-GraphSAGE  } & BoT-IoT   &   \textbf{1.00} \\
  KNN \cite{s21020446}   & BoT-IoT   &   0.99 \\

  XGBoost \cite{electronics9101565}   & BoT-IoT   &   0.97 \\
  
    \hline

\textbf{ Proposed E-GraphSAGE }  & NF-BoT-IoT   &   \textbf{0.81} \\
  Extra Tree Classifier \cite{Sarhan}   & NF-BoT-IoT   &  0.77 \\
    \hline

\textbf{  Proposed E-GraphSAGE }  & ToN-IoT   &   \textbf{0.87} \\
  Extra Tree Classifier \cite{Sarhan}  & ToN-IoT   &   0.87 \\
    \hline

 \textbf{  Proposed E-GraphSAGE } & NF-ToN-IoT   &   \textbf{0.63} \\
  Extra Tree Classifier \cite{Sarhan}   & NF-ToN-IoT   &   0.60 \\
 \hline
\end{tabular}
\label{table:multi_comp}
\end{table}
%TTTTTTTTTTTTTTTTTTTTTTTTTTTTTTTTTTTTTTTTTTTTTTTTTTT

As in the binary classification case, we have provided a comparison of our E-GraphSAGE-based NIDS with the state-of-the-art classifiers. %
Table~\ref{table:multi_comp} shows the average multiclass weighted F1-Scores of E-GraphSAGE compared with the top one or two best reported results in the literature, depending on availability. We observe that E-GraphSAGE outperforms all the state-of-the-art classifiers, with the only exception of the ToN-IoT dataset, where E-GraphSAGE matches the F1-Score as the best reported results.

Overall, we see that E-GraphSAGE at least matches, and in most cases significantly outperforms the state-of-the-art ML-based NIDS approaches, for both binary and multi-class classification.
This is particularly significant, given our stringent comparison methodology, where we picked the best performing classifiers for each of the four individual NIDS datasets, and performed a pairwise comparison with E-GraphSAGE. This is in contrast to a more typical approach in which a classifier is compared with related approaches on one, two, or three common datasets.

%%%%%%%%%%%%%%%%%%%%%%%%%%%%%%%%%%%%%%%%%%%%%%%%
%%%%%%%%%%%%%%%%%%%%%%%%%%%%%%%%%%%%%%%%%%%%%%%%
\section{Conclusions and Future Work}\label{Conclusions}
This paper presents a novel approach to network intrusion detection based on GNNs. For this, we propose E-GraphSAGE which the capturing of edge features as well as the topological pattern of a network flow graph, and is hence able to implement attack flow detection. In this paper, we focus on the application of E-GraphSAGE for IoT network intrusion detection, and more specifically, for the detection of malicious network flows. 
To the best of our knowledge, this represents the first implementation and extensive evaluation of an GNN-based NIDS for IoT using network flow data. 
Our experimental evaluation based on four IoT NIDS benchmark datasets shows that our E-GraphSAGE-based NIDS performs exceptionally well and  overall outperforms the state-of-the-art ML-based classifiers. 
The evaluation results of our initial system demonstrate the potential of a GNN-based approach for network intrusion detection. In the future, we plan to apply neighbourhood sampling techniques to improve the run-time of the proposed E-GraphSAGE model, particularly exploring non-uniform sampling techniques.  Also, it is worth investigating explainable graph neural network algorithms, such as GNNExplainer \cite{ying2019gnnexplainer}, to get more insights about GNN model outputs.  

\ifCLASSOPTIONcaptionsoff
  \newpage
\fi

\printbibliography

% \end{thebibliography}


% \end{IEEEbiography}


% % insert where needed to balance the two columns on the last page with
% %\newpage



%\vfill


\end{document}